\documentclass[final,3p,times,twocolumn]{elsarticle}
\usepackage{hyperref}
\pdfoutput=1
\usepackage{threeparttable,booktabs,multirow}
\usepackage{graphicx}

\journal{Nuclear Instruments and Methods in Physics Research A}

\biboptions{sort&compress}

\begin{document}

\begin{frontmatter}

\title{Combining $\gamma$-ray and particle spectroscopy with SONIC@HORUS}


\author [col]{S.\,G.~Pickstone\corref{corres}}

\ead{pickstone@ikp.uni-koeln.de}
\author[col]{M.~Weinert}
\author[col]{M.~F\"arber}
\author[col]{F.~Heim}
\author[col]{E.~Hoemann}
\author[col]{J.~Mayer}
\author[col]{M.~M\"uscher}
\author[col]{S.~Prill}
\author[col]{P.~Scholz}
\author[col]{M.~Spieker}
\author[col]{V.~Vielmetter}
\author[col]{J.~Wilhelmy}
\author[col]{A.~Zilges}
\cortext[corres]{Corresponding author}


\address[col]{Institute for Nuclear Physics, University of Cologne, Z\"ulpicher Stra{\ss}e 77, D-50937 Cologne, Germany}

\begin{abstract}
The particle spectrometer SONIC for particle-$\gamma$ coincidence measurements was commissioned at the Institute for Nuclear Physics in Cologne, Germany.
SONIC consists of up to 12 silicon $\mathit{\Delta}E$-$E$ telescopes with a total solid angle coverage of 9\,\%, and will complement HORUS, a $\gamma$-ray spectrometer with 14 HPGe detectors.
The combined setup SONIC@HORUS is used to investigate the $\gamma$-decay behaviour of low-spin states up to the neutron separation threshold excited by light-ion inelastic scattering and transfer reactions using beams provided by a 10\,MV FN Tandem accelerator. 
The particle-$\gamma$ coincidence method will be presented using data from a $^{92}$Mo(p,p'$\gamma$) experiment.
In a $^{119}$Sn(d,X) experiment, excellent particle identification has been achieved because of the good energy resolution of the silicon detectors of approximately 20\,keV.
Due to the non-negligible momentum transfer in the reaction, a Doppler correction of the detected $\gamma$-ray energy has to be performed, using the additional information from measuring the ejectile energy and direction.
The high sensitivity of the setup is demonstrated by the results from a $^{94}$Mo(p,p'$\gamma$) experiment, where small $\gamma$-decay branching ratios have been deduced.
\end{abstract}

\begin{keyword}  
\texttt{$\gamma$-ray spectroscopy \sep particle spectroscopy \sep particle-$\gamma$ coincidence technique \sep $\gamma$-decay branching ratios \sep PDR \sep Collective Excitations}
\end{keyword}

\end{frontmatter}

\section{Introduction}
SONIC@HORUS is a newly designed setup for particle-$\gamma$ coincidence measurements at the 10\,MV FN Tandem accelerator at the Institute for Nuclear Physics in Cologne, Germany. It consists of the new particle spectrometer SONIC, which can house up to 12~$\mathit{\Delta}E$-$E$ telescopes for the identification of ejectiles and measurement of their energy; and the existing $\gamma$-ray spectrometer HORUS with its 14 HPGe detectors, which has already been used for the investigation of Nuclear Structure, e.g.\ by measuring fusion-evaporation reactions \cite{Pascu2015,Bernards2010,Karayonchev2017} and capture reactions for Nuclear Astrophysics \cite{LN_NIM, LN_Zr_PLB, AS_As_PRC}.
The identification of the ejectile is necessary to select light-ion scattering and specific transfer reactions, since those are typically not the dominating reaction channels.
This enables a systematic study of a given nucleus with several excitation mechanisms (e.g.\ (two-)neutron transfer and pickup), which gives valuable additional structure information. 
By measuring the energy of the ejectile, nuclear level schemes can be studied in detail, since the excitation energy of the nucleus can be calculated from the ejectile energy.
As presented in \cite{AH_DSAM_NIM}, the knowledge of the complete reaction kinematics also enables a precise and reliable determination of lifetimes using the Doppler-shift attenuation method (DSAM), which has already been applied for several experiments performed at SONIC@HORUS \cite{AH_Ru_PRC, AH_N52_PRC, MS_Diss}.
This paper will present the setup in more detail and will highlight a second main purpose of the setup: the determination of $\gamma$-decay branching ratios, especially for high-lying low-spin states.
These branching ratios probe the overlap of the wave functions of the initial and final states and therefore, are a valuable observable for a comparison with theory. 
Experimentally, the clear assignment of $\gamma$-ray transitions to a certain level can be challenging. However, employing the particle-$\gamma$ coincidence technique --- and thus, the knowledge of the excitation energy of the nucleus --- a safe assignment is possible by applying offline gates to the data as explained in Section~\ref{p-g-coinc}.
One case for which the knowledge of $\gamma$-decay branching ratios is important is the low-energy enhancement of electric dipole strength, commonly denoted by Pygmy Dipole Resonance (PDR) \cite{Savran2013,Bracco2015,Paar2007}. The excitation of this mode has been studied extensively (e.g.\ \cite{JE_Sn_PRL,Savran2006,JE_BaCe_PRC,Pellegri2014,Crespi2015,Crespi2014a,Tamii2011,Schwengner2013a}), mainly in inelastic particle scattering at higher energies and real photon scattering.
However, the $\gamma$-decay branching behaviour	still needs further investigation: Average branching ratios have already been deduced for several nuclei (e.g.\ \cite{Loher2016,Isaak2013,Scheck2013a,Angell2012,Angell2012err,Massarczyk2012,Tonchev2010}), while a comprehensive, state-to-state determination of branching ratios has only been performed in four cases \cite{Scheck2013,Goddard2013,VD_Mo_NPA,Romig2015}. 
With the SONIC@HORUS array, this database will be vastly extended.
In this paper, the experimental setup will be presented in detail in Section~\ref{setup}. To show the power of the particle-$\gamma$ coincidence technique, Section~\ref{p-g-coinc} will be dedicated to gate conditions which can be applied in the analysis.
Since the determination of $\gamma$-decay branching ratios with this method mainly relies on a correct determination of the $\gamma$-ray detection efficiency, an activation measurement to fix the full-energy peak efficiency up to 7\,MeV is presented in Section~\ref{activation}. 
Section~\ref{exp} will report and discuss the experiments performed so far and the performance of the setup:
The clear particle identification is shown in Section~\ref{PID}. During the inelastic scattering process, a non-negligible momentum is transferred to the recoil nucleus and thus, the Doppler-correction demonstrated in Section~\ref{doppler-corr} has been applied. In Section~\ref{94Mo}, $\gamma$-decay branching ratios deduced for $^{94}$Mo will be presented and compared to literature values.
	
\section{Experimental setup}\label{setup}
The beam of light ions is provided by the in-house 10\,MV FN Tandem accelerator at the University of Cologne.
It has a diameter smaller than 3\,mm and a well-defined energy, i.e.\ a typical spread of $\pm$4\,keV for a proton beam of 10.5\,MeV \cite{LN_NIM}).
For the experiments discussed in this paper, the beam energy was chosen between 10\,MeV and 13.5\,MeV. The targets were metallic, self-supporting foils of highly-enriched material with an areal density of 0.3\,mg/cm$^2$ to 0.6\,mg/cm$^2$.

\subsection[g-ray spectrometer HORUS]{$\gamma$-ray spectrometer HORUS}
The $\gamma$-ray spectrometer HORUS (High-efficiency observatory for unique $\gamma$-ray spectroscopy) \cite{Linnemann_Diss} consists of up to 14 HPGe detectors, six of which are actively shielded by anti-Compton BGO shields.
The total full-energy peak efficiency is typically 2\,\% at 1332\,keV and the resolution is of the order of 2\,keV at 6\,kHz and 1332\,keV. The detector positions are on the corners and faces of a cube, which enables the measurement of $\gamma$-$\gamma$ angular correlations as well as particle-$\gamma$ angular correlations.
The particle-$\gamma$ angular correlations measured with the setup are in good agreement with theoretical calculations; details will be discussed in an upcoming publication.
Additionally, the high granularity can be used to determine $\gamma$-ray energy centroid shifts and deduce lifetimes with the DSA method \cite{AH_DSAM_NIM} or to correct the Doppler-shift as explained in Section~\ref{doppler-corr}.
\subsection{Particle spectrometer SONIC}
\begin{figure}[ht]
	\centering
	\includegraphics[width=0.8\columnwidth]{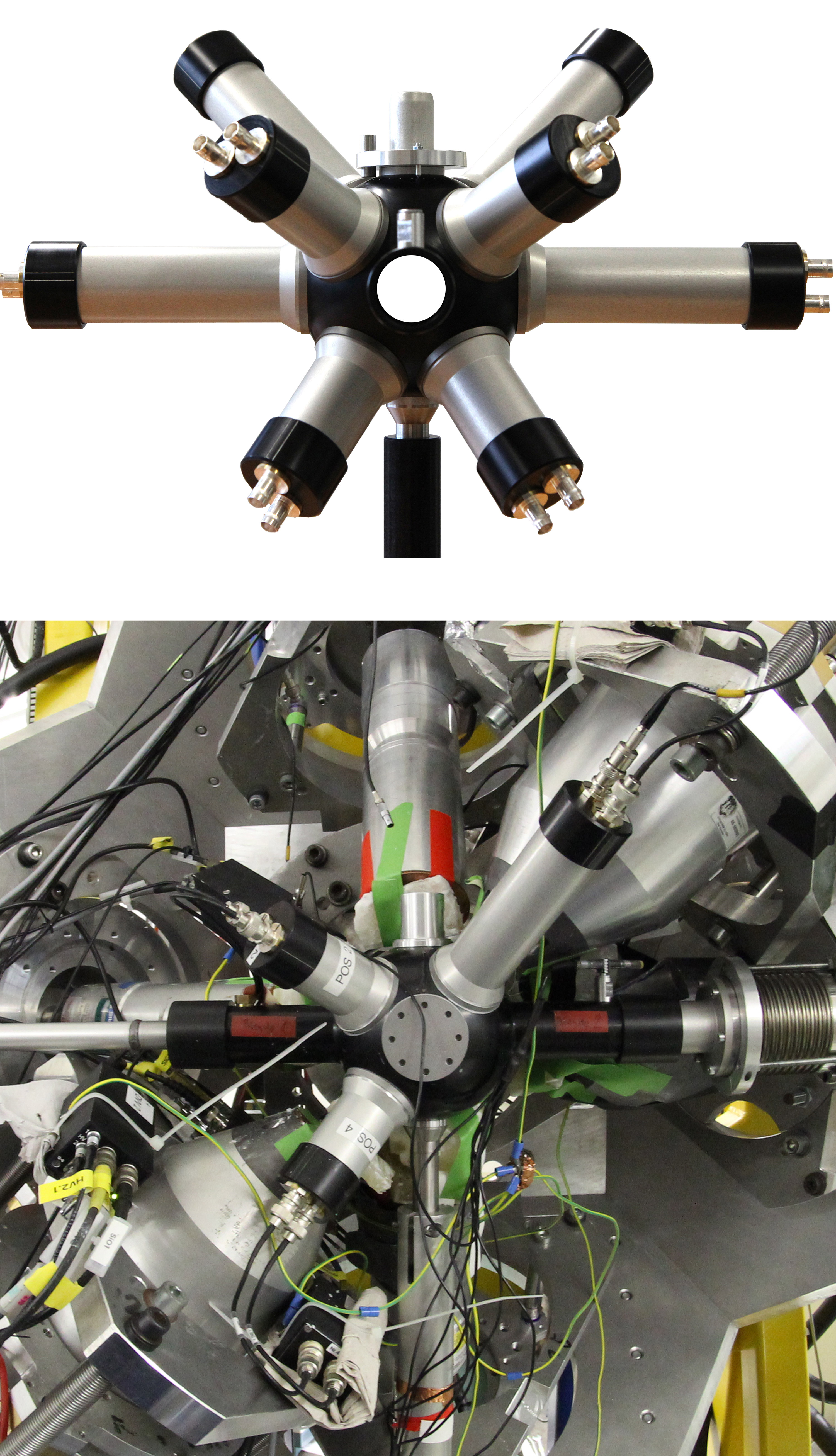}
	\caption{Photograph of SONIC Version I itself (top) and inside the HORUS array (bottom); the outer diameter of the chamber is 12\,cm. In the upper panel, the beam direction is into the plane. The tubes are visible as well as the connectors for the preamplifiers. In the lower panel, the beam goes from left to right. Only the HORUS hemisphere behind the SONIC array is closed and the HPGe detectors and the BGO shields are visible.}\label{pic_sonic1}
\end{figure}

\begin{figure*}
	\includegraphics[width=1.\textwidth]{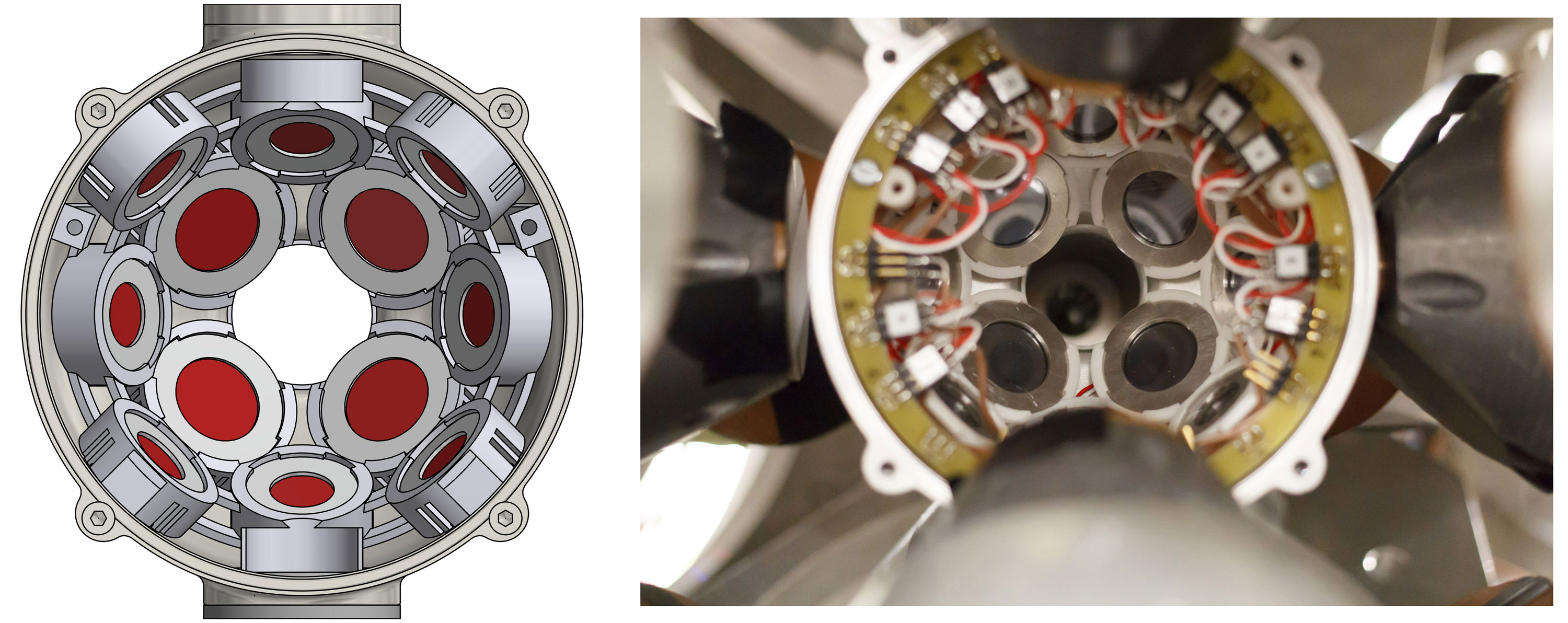}
	\caption{CAD drawing (left) and photograph (right) of SONIC Version III; the diameter of the chamber is 12\,cm. The beam direction is out of the plane. In the CAD drawing, the active areas of the detectors are highlighted. In the photograph, the detectors are in focus, while the readout board on the edge is in the foreground. The lead in front of the BGO shields of the HORUS array is also visible outside the chamber.}\label{pic_sonic3}
\end{figure*}

SONIC (Silicon Identification Chamber) is a new array of up to twelve silicon particle detectors. It can be equipped either with single detectors or with stacked detectors used as telescopes to perform light-ion particle identification (PID) using the $\mathit{\Delta}E$-$E$ technique. There are three versions of the setup, since adjustments were made to better accommodate the individual needs for each experiment.

In the first version, the detectors can be placed at variable distances to the target inside the tube housings (see Fig.~\ref{pic_sonic1}) to customise the count rate of the data acquisition (DAQ).
Since the setup is specifically designed to be embedded in the HORUS array, the tubes are in the gaps between the HPGe detectors to maximize the efficiency of the $\gamma$-ray detection by minimizing the HPGe detector distance to the target while at the same time allowing different silicon detector distances.
The silicon detector preamplifiers are mounted directly onto the tubes, since a short detector-preamplifier distance is paramount to achieve a good energy resolution. As the cross section for elastic scattering is much higher at forward angles, in the standard configuration, four detectors are at backward angles ($\Theta_\textrm{Si}=131^\circ$ and $122^\circ$), while two detectors can be equipped at $\Theta_\textrm{Si}=90^\circ$ and two at forward angles ($\Theta_\textrm{Si}=61^\circ$).

In the second version, the forward detectors are removed, and three additional detectors are positioned by using magnets inside and outside the chamber in between the tubes. The readout of this generation is done via the tubes, which in this configuration cannot be used for particle identification.
However, the solid angle coverage at backward angles is almost doubled as compared to the first version, which allows for the observation of weaker decay channels due to improved statistics.

The third version of SONIC makes use of the advances in rapid prototyping: All detectors are mounted inside the sphere in a plastic frame produced by \emph{selective laser sintering}, which also serves as a cable duct, see Fig.~\ref{pic_sonic3}.
The readout is performed at the base of the setup with preamplifiers for eight channels each.
This third version is now able to house more and bigger detectors, which boosts the solid angle coverage at backward angles to 9\%, more than four times higher than the first version.
The count rate can now be adjusted by tungsten apertures instead of moving the detectors further from the target. This can introduce some background from reactions on the aperture material, but allows for a much more compact setup in order to maximize the $\gamma$-ray detection efficiency. The noise and vacuum performance of this generation is also far superior to the previous generations.
The major design characteristics as well as key features of the different versions are summarised in Table~\ref{tab:summary}.

In all versions, an excellent energy resolution of below 20\,keV for a $^{241}$Am source measurement at $E_\alpha=$ 5486\,keV is achieved by using passivated implanted planar silicon detectors (PIPS from CANBERRA \cite{PIPS_info}) with very thin entrance windows and low leakage currents ($\le$12\,nA for $\mathit{\Delta}E$ and $\le$250\,nA for $E$ detectors at 20\,$^\circ$C).
In-beam, this resolution is typically degraded to approximately 70\,keV due to energy straggling and energy loss in the target as well as kinematic effects. However, the detector energy resolution itself is sufficient to discriminate particles via the $\mathit{\Delta}E$-$E$ technique, which is used for particle identification, see Section \ref{PID}.
Currently, for each telescope, a 0.3\,mm fully depleted $\mathit{\Delta}E$ detector is used in front of a fully depleted $E$ detector with an active thickness of 1\,mm or 1.5\,mm. Nevertheless, the modular mounting design of the third version allows for using different detectors depending on the experimental needs.

Due to the thin entrance window, the detectors must be shielded from $\delta$-electrons, because they lead to a much worse energy resolution for the ejectile due to summing effects. Using high voltage of up to several kV to deflect the electrons proved to be ineffective, so now a thin composite foil (made of polyethylene terephthalate and modified acrylate) with a thickness of 5\,$\mu$m is used in front of all the detectors, 
which effectively shields the electrons without influencing the ejectile detection.

\begin{table}
\caption{Summary of the three versions of the SONIC array. The efficiencies given here are for the single $E$ configuration for comparison. $\Theta_\textrm{Si}$ is the angle relative to the beam axis for a given detector group. $r_{det}$ gives the typical distance of the detector to the target, and $A_{det}$ its active area. $\Omega$ is the solid angle coverage relative to the whole sphere. $d$ denotes the thickness of the detectors.
}\label{tab:summary}

\begin{tabular}{llll}
	\toprule
	version				& I 			& II					&III  \\
	\midrule
\multirow{2}{*}{mount}		& \multirow{2}{*}{tubes} 			& tubes \& 			& plastic  		\\
							&									& magnets			& frame		\vspace{5pt}\\
	\multirow{4}{60pt}{\#detectors @~$\Theta_\mathrm{Si}~(^\circ)$}	& 2 @ 131		& 3 @ 122 & 4 @ 145\\           
	 						& 2 @  122		& 4 @ 114			& 4 @ 123 	\\           
	 						& 2 @  90  		&	 				& 4 @ 107  	\\           
							& 2 @ 61 		& 					&			\vspace{5pt}\\
	
	$r_{det}$ (mm)			& 45 -- 100		& 45 -- 100 			&45				\\
	$A_{det}$ (mm$^2$)			& 150			& 150				& 150, 300\\
	$\Omega / 4\pi$ 		& 4\%			& 4\%  				& 9\% 		\\           
$d_{\mathit{\Delta E}}$ (mm)	& 0.3		& 0.3 				& 0.3 	\\
$d_{E}$ (mm)					& 1.5       & 1.5   			&1.0, 1.5 \\
\bottomrule
	\end{tabular}
\end{table}

\section[Particle-g coincidence analysis]{Particle-$\gamma$ coincidence analysis}\label{p-g-coinc}
\begin{figure*}
	\includegraphics[width=1.0\textwidth]{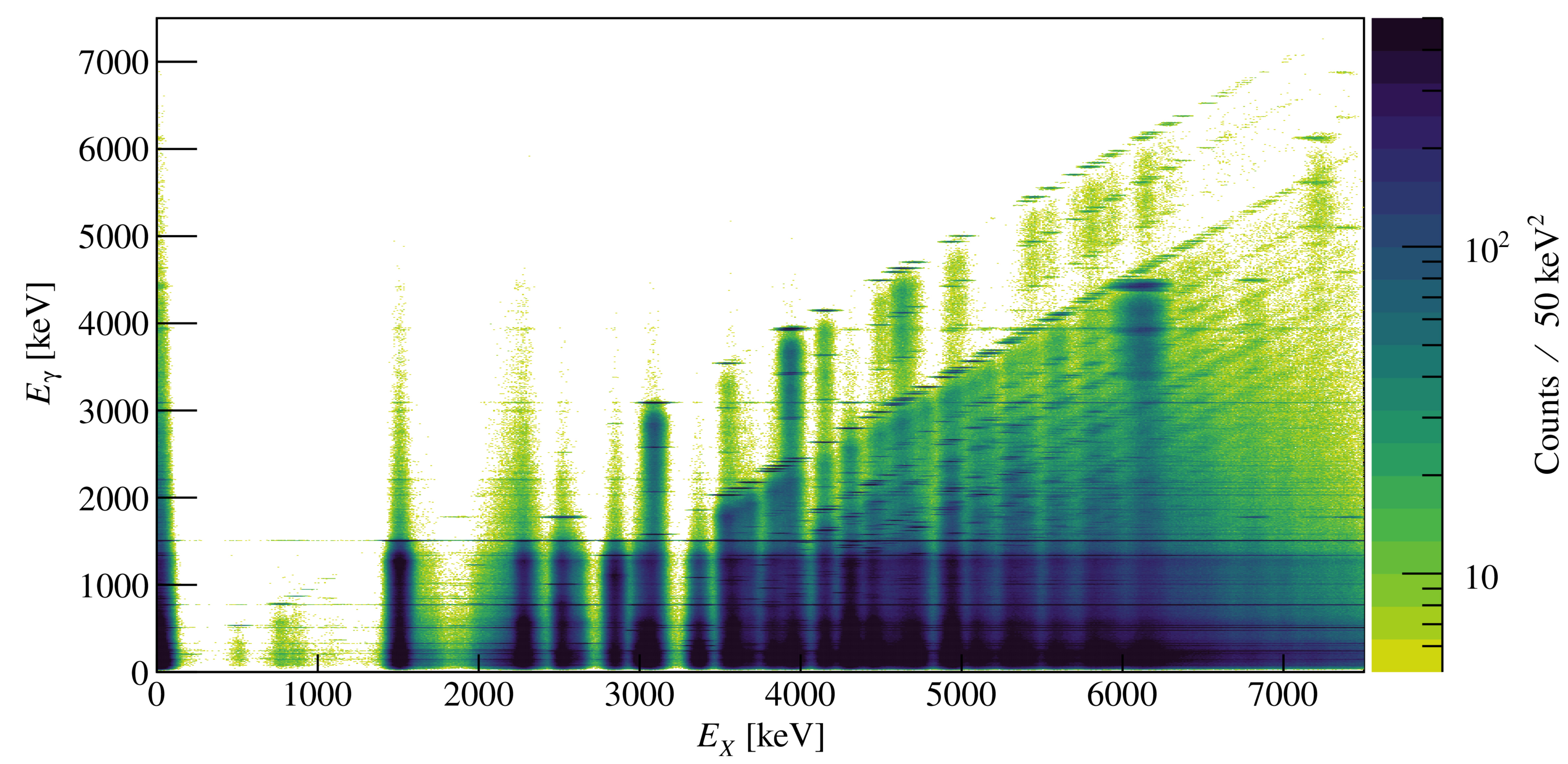}
	\caption{p-$\gamma$ coincidence matrix of the $\gamma$-ray energy vs the excitation energy for the $^{92}$Mo(p,p'$\gamma$) experiment at $E_p$=10.5\,MeV. Columns correspond to the $excitation$ of a certain level in the target nucleus, while rows correspond to a specific $de-exciting$ $\gamma$-ray energy. The diagonal structures correspond to the decay $to$ a certain level with an energy $E_\mathrm{level}$, as they fulfil the condition $E_X=E_\gamma + E_\mathrm{level}$. They are most prominent at higher $\gamma$-ray and excitation energies. The highest diagonal is the ground-state diagonal (followed by the weaker single-escape and double-escape diagonal, which are part of the detector response of the ground-state diagonal). The diagonal shifted downwards by 1510\,keV = $E(2^+_1)$ shows decays to the first excited state in $^{92}$Mo, while diagonals of transitions to higher-lying states are also clearly visible. More details are also given in Section~\ref{p-g-coinc}.}\label{fig:p-g-matrix}
\end{figure*}
\begin{figure}
	\includegraphics[width=1.0\columnwidth]{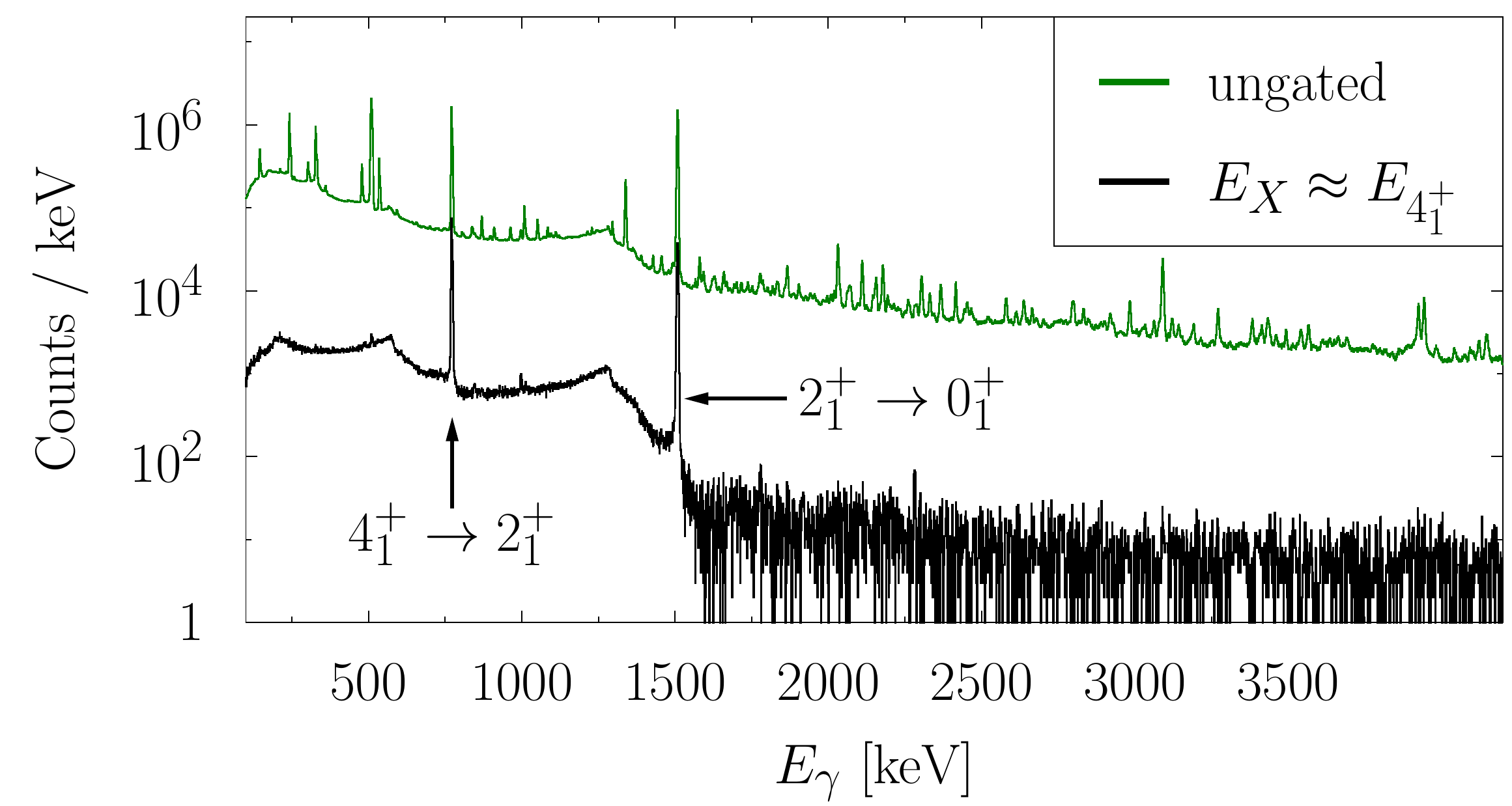}
	\caption{$\gamma$-ray spectrum of the $^{92}$Mo(p,p'$\gamma$) experiment without a gate on a specific excitation energy (upper spectrum) and with a gate on the excitation of the $4^+_1$ (lower spectrum). Note the logarithmic scale, which emphasizes the tremendous improvement of the peak-to-background ratio from the additional excitation energy information from the particle detectors.
	}\label{fig:x-gate}
\end{figure}

The data from SONIC@HORUS are recorded with the DGF-4C Rev.\ F modules from XIA \cite{XIA_NIM} and stored as listmode files for multiplicity $\ge$2 events. All selection processes are then performed offline to retain as much information as possible for the analysis.
The time information of the detectors is used to define a prompt peak in the time spectrum and perform a background subtraction, if necessary.
The particle detectors are calibrated to excitation energy and a particle-$\gamma$ matrix is constructed, see Fig.~\ref{fig:p-g-matrix}. In this matrix, the different energy resolutions of the particle and $\gamma$-ray detectors are clearly visible. 

For analysis, a certain excitation energy (range) can be required (`gating') and projecting the resulting spectrum on the $\gamma$-ray energy axis already reveals the complete decay pattern and its detector response of a given state with the superior energy resolution of the HPGe detectors. Figure \ref{fig:x-gate} shows the $\gamma$-ray spectrum with such a gate on the excitation energy of the $4^+_1$ in $^{92}$Mo compared to a $\gamma$-ray spectrum for events where any energy was deposited in the particle detectors. The strongly improved peak-to-background ratio is clearly visible as well as the complete elimination of other cascades.
Additionally, for states where decays to only a small number of final states are probable (e.g.\ due to the favourably small $\Delta L$ in electromagnetic transitions), another approach can be used:

For direct decay back to the ground state, the excitation energy $E_X$ of the initial level matches the de-excitation energy $E_\gamma$. Similarly, spectra with transitions only to a given level at energy $E_f$ can be selected by gating on $E_\gamma + E_f \approx E_X$.

By comparing these spectra, the calculation of $\gamma$-decay branching ratios is rather straightforward and independent of the excitation of the specific initial level.
Particle-$\gamma$ angular correlations can be neglected for the results presented in this paper, so
only the efficiency remains to be determined for the deduction of branching ratios, which will be discussed in the following section.

\section{High-energy efficiency determination}\label{activation}
\begin{figure}[ht]
	\centering
	\includegraphics[width=1\columnwidth]{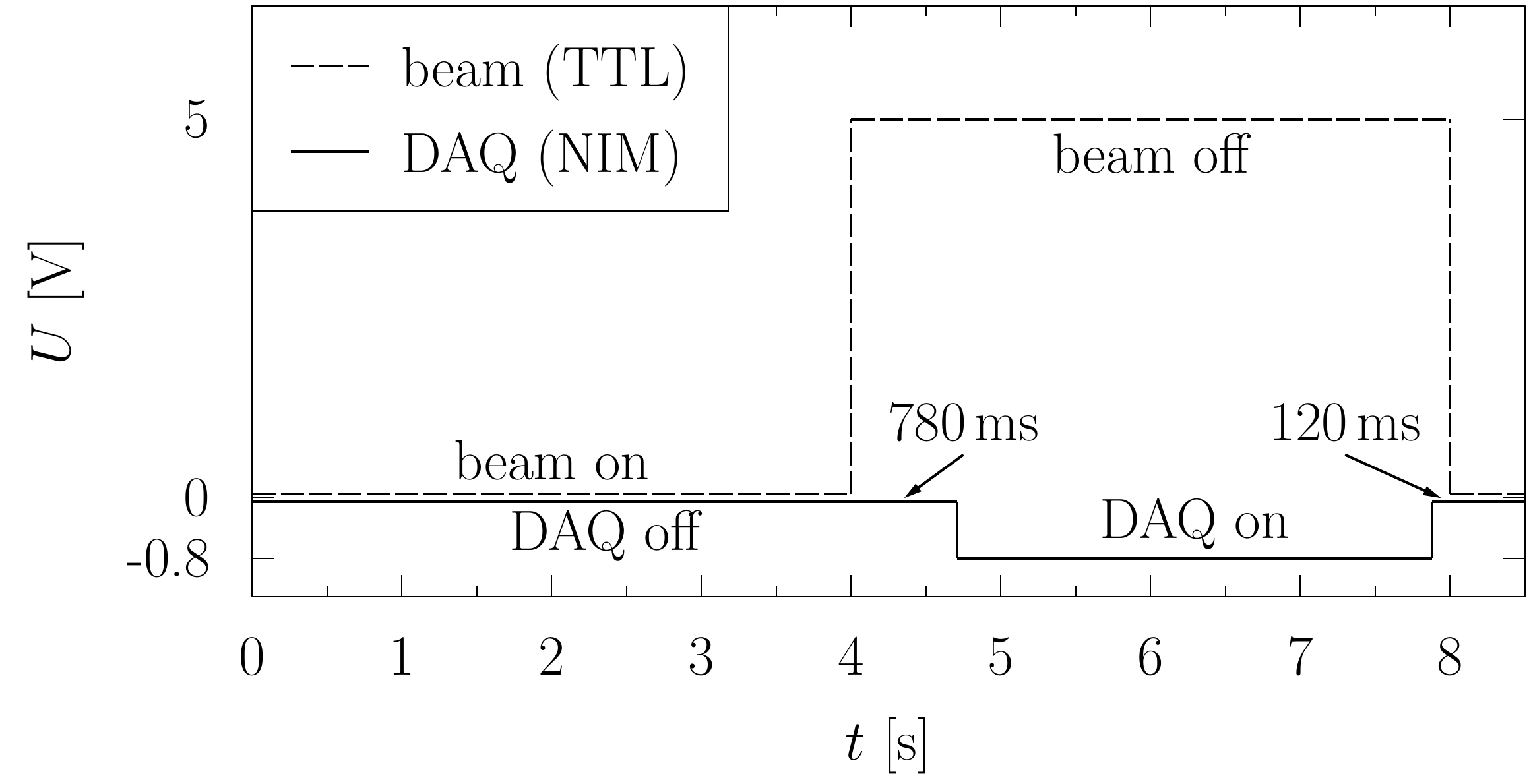}
	\includegraphics[width=1\columnwidth]{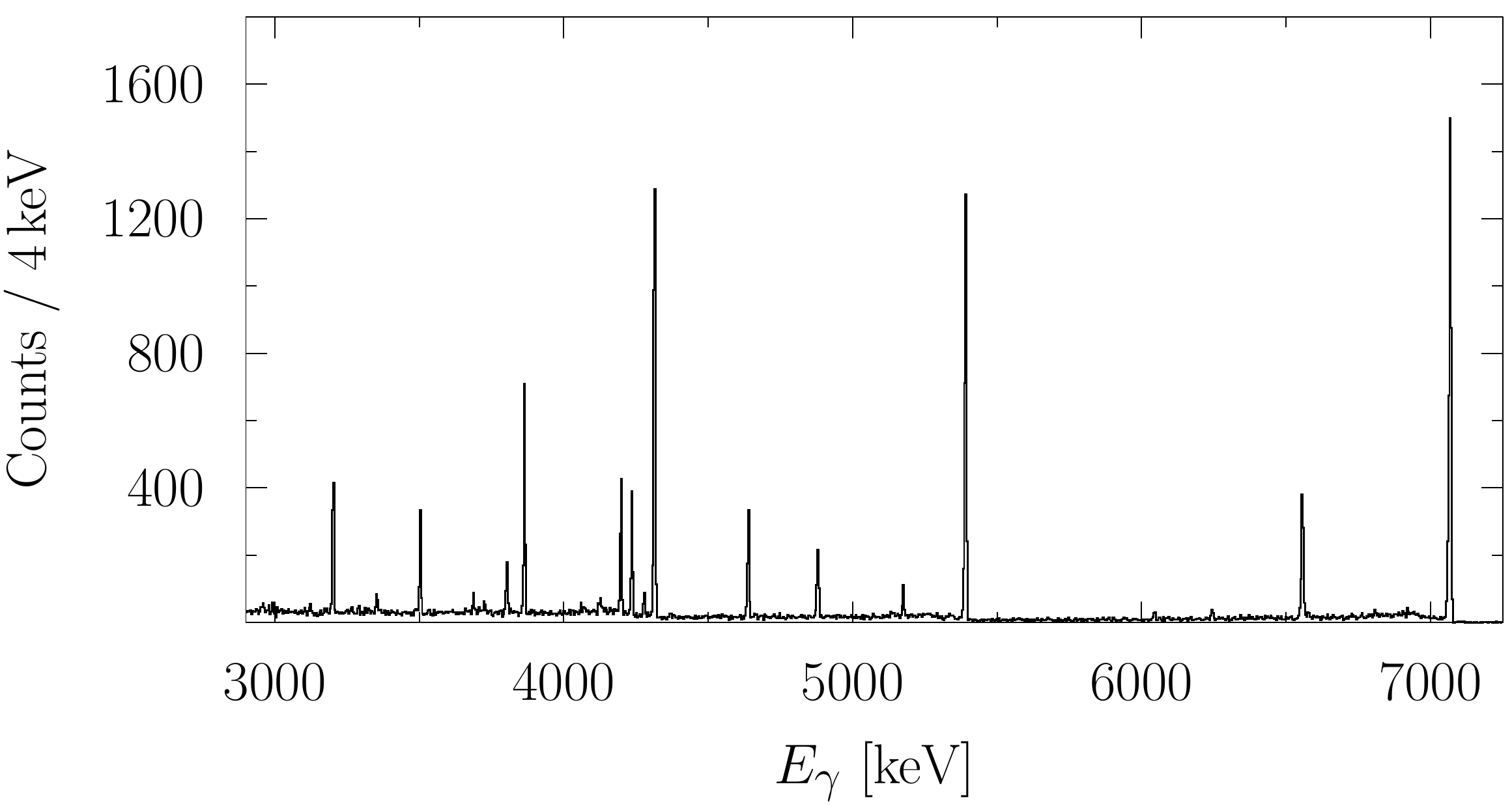}
	\caption{Timing diagram and $\gamma$-ray spectrum for the $^{24}$Mg(p,n)$^{24}$Al activation and the decay of $^{24}$Al. In the upper panel, the timing of the beam and the DAQ are shown.
	The 780\,ms gap after activation suppresses contributions from the 131\,ms isomer of $^{24}$Al.
	In the lower panel, the high-energy part of the $\gamma$-ray spectrum of one detector for the $^{24}$Al decay is shown. All visible peaks are decay radiation from $^{24}$Al and detector response. Note the low background thanks to the offline condition.} \label{activation-spec}\label{activation-timing}
\end{figure}

For low energies (below 3.5\,MeV), the efficiency can be easily determined with standard calibration sources, e.g.\ $^{226}$Ra and $^{56}$Co. However, for higher energies, no long-lived calibration source is available and therefore, it has to be produced on-site.
One reaction, which had already been used to measure high-energy efficiencies, is the $^{24}$Mg(p,n)$^{24}$Al activation (see, e.g.\ \cite{Wilhelm1996}), which was performed at a proton energy of 18\,MeV. The half-life of $^{24}$Al is 2\,s \cite{NDS024}, so a pulsed beam with a bunch length of 4\,s was chosen for activation.
After a short break of 0.8\,s to avoid a contribution of $\gamma$-rays from the isomeric decay of $^{24}$Al (with a half-life of $T_{1/2}=131$\,ms \cite{NDS024}), data were recorded for 3.1\,s. The timing pattern and the resulting spectrum are shown in Fig.~\ref{activation-spec}. From this spectrum, a reliable high-energy efficiency up to 7\,MeV can be extracted by interpolation rather than by relying on simulations or extrapolation alone. 

\section{Experimental results}\label{exp}
\begin{figure}[ht]
	\centering
	\includegraphics[width=1\columnwidth]{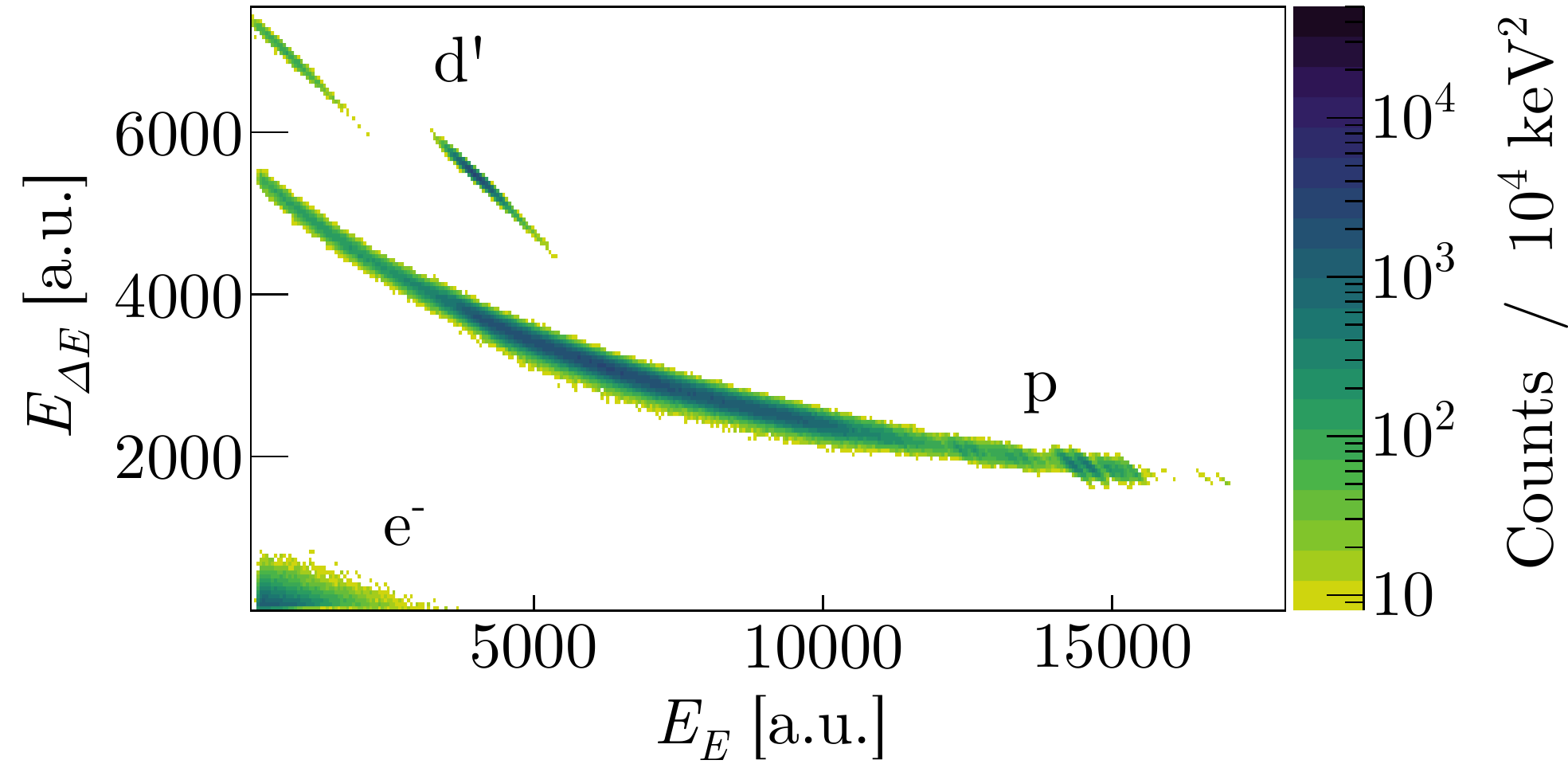}
	\includegraphics[width=1\columnwidth]{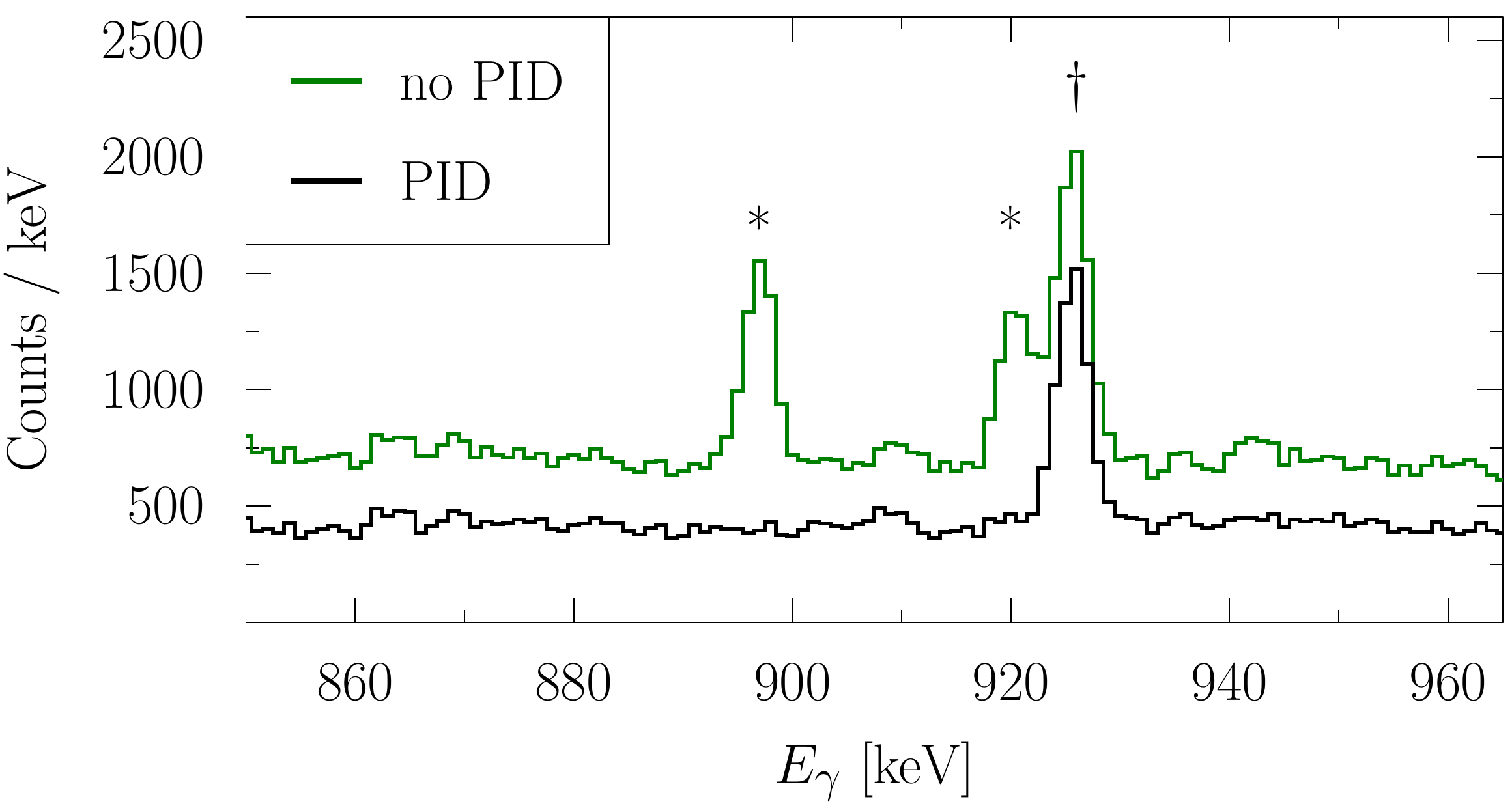}
	\caption{PID in the $^{119}$Sn(d,X) experiment. In the upper panel, the energy detected in the first detector ($E_{\Delta E}$) is plotted vs the energy deposition in the second detector ($E_E$). The types of particles are marked accordingly. Note the clear discriminatio as well as the very low background. The structures visible in the proton area at around $E_E=15000$\,a.u. correspond to the excitation of levels in $^{120}$Sn. In the lower panel, the spectrum with gate on protons as ejectiles (lower spectrum) is compared to the ungated data (upper spectrum). The suppression of the $\gamma$-ray transitions in $^{119}$Sn (*) is evident, while the $\gamma$-ray transition in $^{120}$Sn ($\dagger$) remains. For further details, see Section \ref{PID}.	}\label{pid-spec}
\end{figure}
\subsection[Particle Identification in a 119Sn(d,pg)120Sn experiment]{Particle Identification in a $^{119}$Sn(d,p$\gamma$)$^{120}$Sn experiment}\label{PID} 
The particle identification capabilities presented in this paper were recorded with a deuteron beam of 10\,MeV impinging on a 93\% enriched $^{119}$Sn target with a thickness of 0.4\,mg/cm$^2$. The Q-value for the $^{119}$Sn(d,p)$^{120}$Sn reaction is 6.9\,MeV \cite{AME2016_2}. The first version of SONIC was used, equipped with four $\mathit{\Delta}E$-$E$~telescopes at backward angles, consisting of detectors of 0.3\,mm and 1.5\,mm thickness.
The corresponding $\mathit{\Delta}E$-$E$~matrix is shown in the upper panel of Fig.~\ref{pid-spec}.
Protons from the transfer reaction, deuterons from (in-)elastic scattering and electrons are clearly distinguishable.
In the lower panel of Fig.~\ref{pid-spec}, the $\gamma$-ray spectra with and without particle identification are compared: The suppression of the $\gamma$-ray transitions in $^{119}$Sn at 897\,keV and 921\,keV \cite{NDS119} is evident, while the $\gamma$-ray transition in $^{120}$Sn at 926\,keV \cite{NDS120} remains.

\subsection[Doppler-correction in a 60Ni(p,p'g)60Ni experiment]{Doppler-correction in a $^{60}$Ni(p,p'$\gamma$)$^{60}$Ni experiment}\label{doppler-corr}

After the reaction, the nucleus is moving at a certain velocity relative to the $\gamma$-ray detector due to the momentum transferred in the collision.
The detected $\gamma$-ray energy is then Doppler-shifted on an event-by-event basis depending on both the ejectile and $\gamma$-ray angles and energies.
This shift can be used to extract lifetimes via the Doppler-Shift Attenuation Method as explained in \cite{AH_DSAM_NIM}, which has already been applied to several experiments \cite{AH_Ru_PRC, AH_N52_PRC, MS_Diss}.
 
For the experiments presented here, the Doppler-shift is $corrected$ on an event-by-event bases instead. Following the notation used in \cite{AH_DSAM_NIM}, the spectra can be corrected by the following formula:
\[
E_\gamma^0 =
\frac{E_\gamma(\Theta)}{(1+F(\tau)\frac{v_0}{c}\cos\Theta)} \approx
\frac{E_\gamma(\Theta)}{(1+\frac{v_0}{c}\cos\Theta)}
\]
$E_\gamma(\Theta)$ is  the detected $\gamma$-ray energy, $E_\gamma^0$ is the corrected, true $\gamma$-ray energy and $v_0$ is the initial recoil velocity, which is calculated from the detected ejectile energy and direction \cite{AH_DSAM_NIM}. $\Theta$ is the relative angle of the recoiling nucleus to the de-exciting $\gamma$-ray, calculated also from the reaction kinematics. 
For the correction, the attenuation factor $F(\tau)$ is assumed to be 1, which is valid for short lifetimes (shorter than $\sim$\,10\,fs --- as is the case for the PDR states (see, e.g.\ \cite{Scheck2013})) and thin targets without a stopper layer, because most $\gamma$-rays are then emitted from nuclei close to the initial recoil velocity $v_0$.

\begin{figure}	
	\includegraphics[width=1.0\columnwidth]{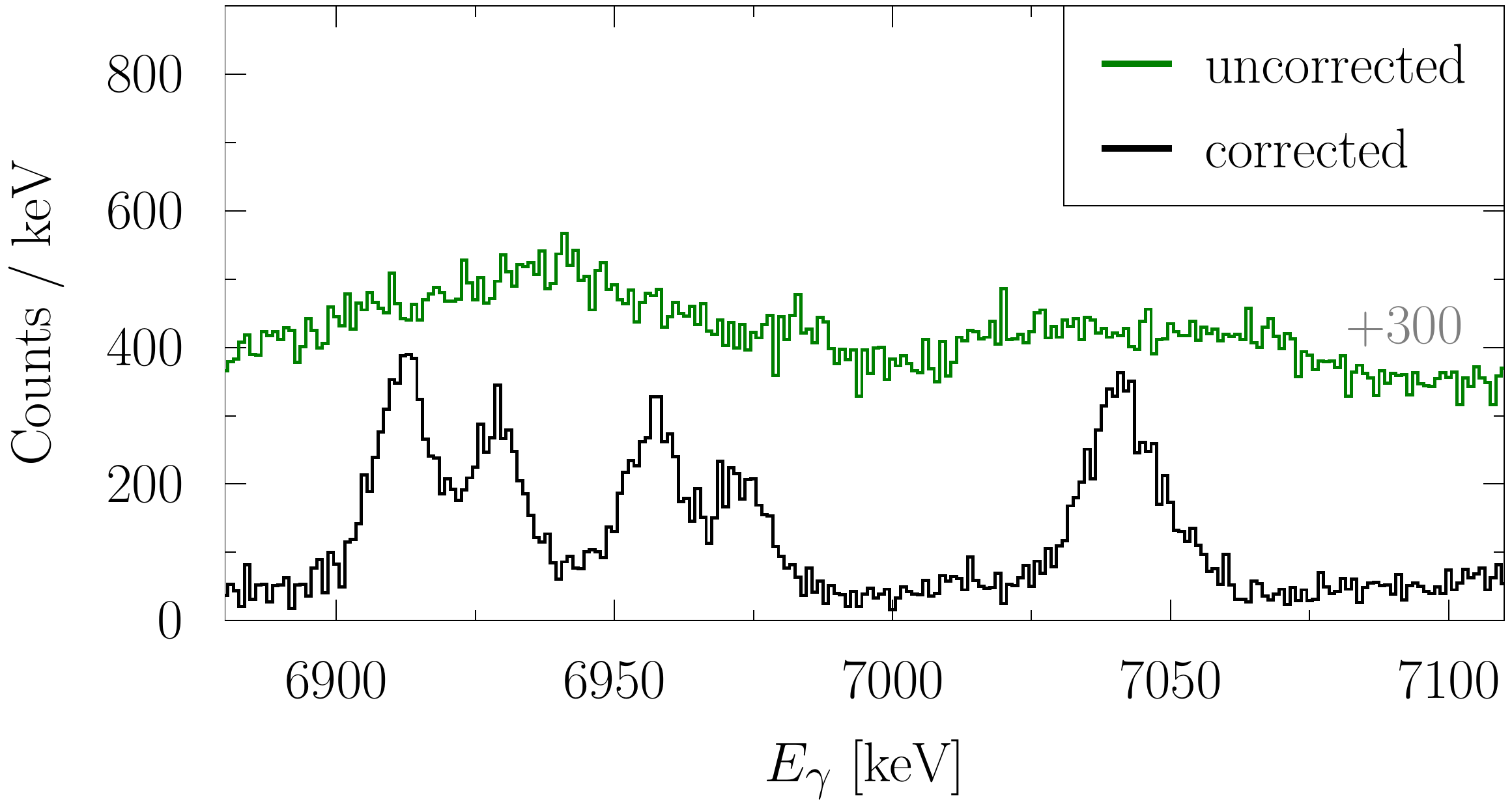}
	\caption{Doppler-correction in the $^{60}$Ni(p,p'$\gamma$) experiment. The corrected spectrum (lower spectrum) clearly shows transitions in the high-energy region, while the uncorrected spectrum (upper spectrum, shifted by 300 counts) does not allow for a clean analysis. The FWHM is improved from 50\,keV (uncorrected) to 15\,keV (corrected) at these energies.}\label{comp_doppler}	
\end{figure}

The effect of the correction on the high-energy $\gamma$-ray spectrum in the HORUS detectors is shown in Fig.~\ref{comp_doppler} for data recorded with the third version of SONIC in a  $^{60}$Ni(p,p'$\gamma$)$^{60}$Ni experiment at $E_p$~=~15\,MeV. The areal density of the target was 0.35\,mg/cm$^2$, and the enrichment was 99\%. The figure shows a summed spectrum over all active detectors of the HORUS array in this experiment, with a gate on ground-state transitions. The FWHM, e.g.\ around 7\,MeV, is 50\,keV for the uncorrected spectrum and is greatly improved to 15\,keV after correction due to the improved single detector resolution as well as the improved calibration alignment of the detector array.
A strong improvement of the $\gamma$-ray energy resolution was also observed for experiments with heavier target nuclei and lower beam energies, so the Doppler-correction was applied to all experiments presented here.

\subsection[Gamma-decay branching ratios for 94Mo]{$\gamma$-decay branching ratios for $^{94}$Mo}\label{94Mo}
\begin{table}[ht]
	\begin{threeparttable}
		\caption{Comparison of the $\gamma$-decay branching ratios measured in the $^{94}$Mo(p,p'$\gamma$) experiment at $E_p = 13.5$\,MeV ($BR_{meas}$) to literature values taken from \cite{NDS094} ($BR_{lit}$), normalised to the strongest decays. The second version of SONIC was used in this experiment. For $BR_{meas}$, only statistical uncertainties are given.}\label{tab:decays}
		\begin{tabular}{cccccc}
			\toprule
			$E_i $	& $J^\pi_i$	& $E_{f}$ 		& $J^\pi_f$	& $BR_{meas}$	& $BR_{lit}$\\
			(keV)	& 			& (keV) 		& 			& (\%)			& (\%)	\\
			\midrule
			1864	& $2^+$		& 871			& $2^+$		& 100.0(9)		& 100.0(8)	\\
					&			& 0				& $0^+$		& 12.8(3)		& 8.9(11)	\\
			
			2067	& $2^+$		& 871			& $2^+$		& 100.0(11)		& 100.0(7)	\\
					&			& 0				& $0^+$		& 14.0(4)		& 15.1(7)	\\
			
			2393	& $2^+$		& 871			& $2^+$		& 100(2)		& 100(2)	\\
					&			& 0				& $0^+$		& 12.5(6)		& 11.1(2)	\\	
			
			2740	& $1^+$		& 1864			& $2^+$		& $\leq$60		& 24(1)		\\
					&			& 1742			& $0^+$		& $\leq$5		& 4.4(1)	\\
					&			& 871			& $2^+$		& 100(4)		& 100(2)	\\
					&			& 0				& $0^+$		& 59(3)			& 65(1)		\\	
			
			2870	& $2^+$		& 2393			& $2^+$		& 13(1)			& -			\\
					& 			& 2067			& $2^+$		& 11(1)			& 26(2)		\\
					&			& 1864			& $2^+$		& 100(2)		& 100(4)	\\
					&			& 1742			& $0^+$		& 12(1)			& -			\\
					&			& 871			& $2^+$		& 11(1)			& 13.1(6)	\\
					&			& 0				& $0^+$		& 19(1)			& 17.3(5)	\\			
			
			3261	& $1^-$ 	& 2067			& $2^+$ 	& 16.3(7) 		& -			\\
					& 			& 1742			& $0^+$ 	& 19.2(8)		& -			\\
					& 			& 871 			& $2^+$		& 12.3(7)		& 67(17)?$^\dagger$	\\
					&			& 0 			& $0^+$		& 100(2)		& 100(17)	\\
			\bottomrule
		\end{tabular}
		\begin{tablenotes}
			\item[$\dagger$] \footnotesize{Reported only in \cite{Sugiyama1976}. Placement in the level scheme questionable there, since the ground-state transition of the $2^+_4$ is very close in $\gamma$-ray energy.}
		\end{tablenotes}	
	\end{threeparttable}
\end{table}

As mentioned in the introduction, the comprehensive determination of $\gamma$-decay branching ratios is one main purpose of the SONIC@HORUS array. A comparison of these branching ratios ($BR$) with existing literature values is shown in Table~\ref{tab:decays}.

The overall agreement is very good and shows the capabilities of the SONIC@HORUS array. For some states, new branching ratios have been determined. The comprehensive determination of the $\gamma$-decay behaviour of the $1^-$ state at 3261\,keV serves as an example of how the presented setup can extend the knowledge of smaller decay branches. The detailed investigation of higher-lying states and their decay branching ratios will be discussed in upcoming publications.

\section{Conclusions}
The SONIC@HORUS setup was presented in detail, including its different versions. The particle-$\gamma$ coincidence method was shortly summarised, focussing on the aspects relevant for the analysis of experiments performed with this array. The high-energy efficiency measurement which has been re-established in Cologne was also presented.
Experimental results from the first physics experiments were given, showing both the particle identification capability as well as the big improvement of the $\gamma$-ray spectra by applying the Doppler-correction.
$\gamma$-decay branching ratios for an inelastic scattering experiment on $^{94}$Mo were presented, showing the good agreement with existing data as well as newly observed transitions.
In upcoming publications, the discussion of the $\gamma$-decay behaviour for all experiments presented here will be discussed in more detail, focussing on the high-energy dipole response. 

\section*{Acknowledgements}
The authors would like to thank both the accelerator and workshop staff at the Institute for Nuclear Physics as well as J.~Endres and A.~Schreckling. This work is supported by the Deutsche For\-schungs\-ge\-mein\-schaft under contract ZI-510/7-1. S.\,P.\ and J.\,W.\ are supported by the Bonn-Cologne Graduate School of Physics and Astronomy.

\bibliography{library_2017-08-02}
\end{document}